\documentclass[sigconf]{acmart}

\AtBeginDocument{%
  }

\copyrightyear{2026}
\acmYear{2026}
\setcopyright{cc}
\setcctype{by}
\acmConference[IDE '26]{3rd International Workshop on Integrated Development Environments }{April 12--18, 2026}{Rio de Janeiro, Brazil}
\acmBooktitle{3rd International Workshop on Integrated Development Environments (IDE '26), April 12--18, 2026, Rio de Janeiro, Brazil}
\acmPrice{}
\acmDOI{10.1145/3786151.3788607}
\acmISBN{979-8-4007-2384-1/2026/04}

\usepackage{url}            
\makeatletter
\def\url@leostyle{%
  \@ifundefined{selectfont}{\def\UrlFont{\sf}}{\def\UrlFont{\small\sffamily}}}
\makeatother
\usepackage{listings}
\lstdefinestyle{idstyle}{
  basicstyle=\ttfamily\small,
  breaklines=true,
  columns=fullflexible
}
\newcommand{\idstyle}{\ttfamily\bfseries} 

\newcommand{\codefontfamily}{\ttfamily}
\newcommand{\myCommentStyle}[1]{{\small\codefontfamily\color{gray!100!white} #1}}
\newcommand{\myStringStyle}[1]{{\small\codefontfamily\color{violet!100!black} #1}}

\newcommand{\myKeywordStyle}[1]{{\small\codefontfamily\color{green!70!black} #1}}

\begin{document}

\title{It's Alive! What a Live Object Environment Changes in Software Engineering Practice}


\author{Juli\'an  Grigera}
\email{juliang@lifia.info.unlp.edu.ar}
\orcid{0000-0002-7962-4312}
\affiliation{%
  \institution{LIFIA, Fac. de Inform\'atica, Univ. Nac. de La Plata}
  \city{La Plata}\country{Argentina}
  }
\affiliation{%
  \institution{CONICET}
  \city{Buenos Aires}\country{Argentina}
  }

\author{Steven Costiou}
\email{steven.costiou@inria.fr}
\orcid{0000-0003-2787-5432}
\affiliation{%
  \institution{Inria}
  \city{Lille}
  \country{France}
}
  
\author{Juan Cruz Gardey}
\email{jcgardey@lifia.info.unlp.edu.ar}
\orcid{0000-0002-1765-8189}
\affiliation{%
  \institution{LIFIA, Fac. de Inform\'atica, Univ. Nac. de La Plata}
  \city{La Plata}\country{Argentina}
  }

\author{St\'ephane Ducasse}
\email{stephane.ducasse@inria.fr}
\orcid{0000-0001-6070-6599}
\affiliation{%
  \institution{Inria}
  \city{Lille}
  \country{France}
}

\renewcommand{\shortauthors}{Grigera et al.}

\begin{abstract}
Tools shape our mind. 
This is why it is important to have extensible and flexible tools for developers to adapt to their needs. 
Reasoning about programs in the abstract —by imagining what objects should look like— can make it harder to grasp the underlying model.
In Smalltalk environments like Pharo, developers work closely with their objects, gaining immediate feedback - not guessing how they will look like but directly interacting with them.
This article presents some tools developers use in Pharo: Inspector custom views for defining specific views and navigation for objects, Microcommits for reverting changes without the need to commit and pull, Xtreme TDD that allows developers to code in the debugger, On the Fly Rewriting Deprecations that support API evolution through automated rewriting of deprecated calls, and Object-Centric Breakpoints - when a problem cannot be efficiently solved with a dummy trace, developers can use break points that will only halt for a given instance.
By showcasing these features that evolved alongside Smalltalk, we invite reflection on how other IDEs could rethink some of their features and improve developers' workflows.
\end{abstract}


\begin{teaserfigure}
  \includegraphics[width=\textwidth]{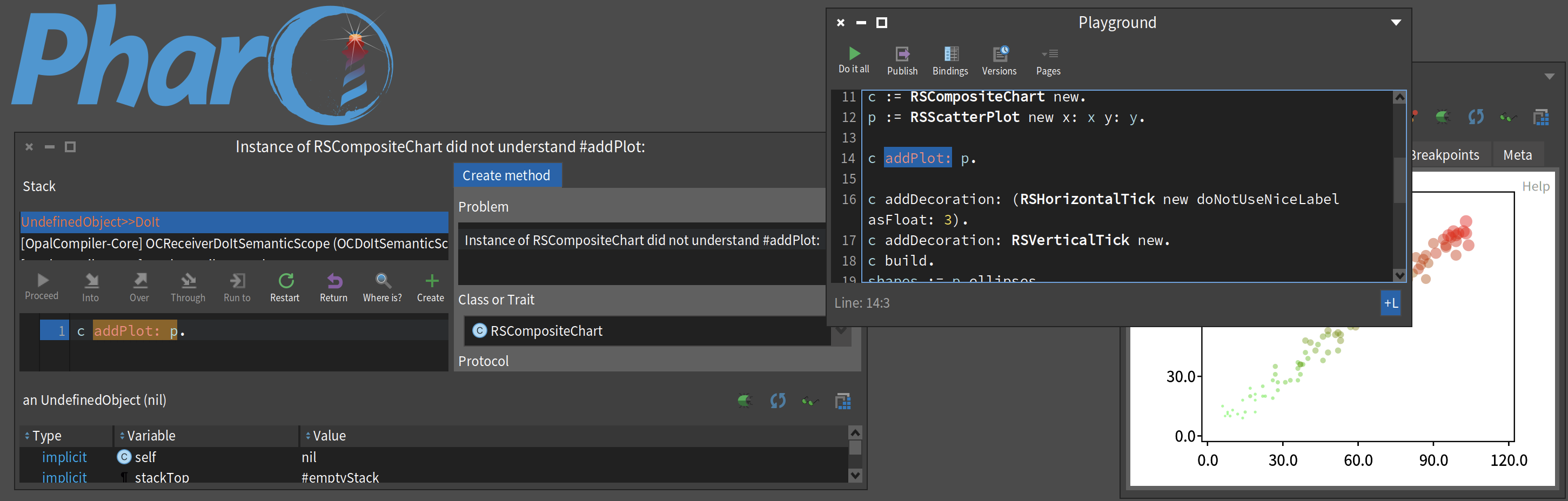}
  \caption{Pharo: an IDE designed for live programming and XTDD.}
  \Description{A screenshot of the Pharo Environment}
  \label{fig:teaser}
\end{teaserfigure}

\received{27 October 2025}
\received[accepted]{27 November 2025}

\maketitle


\section{Introduction}

Most modern IDEs did not redefine how we build software — they inherited it. 
They arrived after the languages they support, and wrapped the existing compile–run–debug pipeline that came from the command-line era. 
As a result, the coding process is still largely based on this pipeline.
This involves working around files (which is what compilers expect), launching a separate process to run a program, or debugging in an external mode rather than part of execution.
Most tools aimed at improving this process are bolt-ons that observe the runtime from the outside. 
Even though today’s IDEs have come a long way, they still promote a linear workflow of edit > build > run > debug, shaping the software engineering process into discrete phases with little continuity of state.

Smalltalk environments — and Pharo in particular — were developed with a different concept \cite{Gold84a, Blac09a}. 
They were co-designed with the language as one live system: the IDE itself is the running program.
Code exists as objects in memory rather than text in files, debugging happens inside the execution context, and developers can change methods and immediately resume execution. 
Tools such as browsers and inspectors share the same object space as the application, blurring the line between “editing”, “running” and “analyzing”. 
As a result, development emphasizes continuous interaction with the running system.

The point is not about Pharo's unique features, but a different mental model of software creation \cite{Kube15a,Kube18a,Kube19a}.
While some practices, such as Extreme TDD, could in principle be adopted in other environments, Pharo demonstrates how tight integration between tools can make such practices routine rather than exceptional.
By contrasting these two lineages — the file-based, phase-driven tradition of mainstream IDEs and the live, reflective architecture of Pharo \cite{Thoma24a} — we can ask how IDE design shapes software engineering practice, and what insights from Pharo might inform the next generation of development environments.

The following sections illustrate 3 scenarios, each highlighting a specific feature of Pharo as an IDE, which could help to reflect on how traditional IDEs could support a different development process inspired by live environments.
To ground these ideas, we use as a running example a simple logistics system that manages packages, depots, trucks, and delivery routes.

\section{Scenario: \textit{Debugger-Driven Development}}
In Pharo, the debugger functions as a fundamental development interface, supporting activities beyond diagnosing errors.
Users can write or modify methods directly in the debugger, and then resume execution without restarting the system. 
This enables \textit{Debugger-Driven Development} \textit{(DDD)} which is not commonly supported in the same form by mainstream IDEs.
 \textit{Debugger-Driven Development} is part of Xtreme TDD (where developers write failing tests and they code in the debugger). 
The design of the Pharo debugger focuses on maintaining a streamlined debugging workflow that helps developers concentrate on a single tool for exploring runtime values and behavior.

To illustrate the dynamic nature of Pharo's debugger, we will describe a scenario in which a developer codes a new feature using \textit{DDD}. 
While working on the logistics system, the developer needs to add a new feature: packages located within a depot’s service area should be automatically scheduled for delivery. 
This extension is implemented using \textit{DDD}, with the debugger driving the development process.

\begin{figure}[H]
\includegraphics[width=0.5\linewidth]{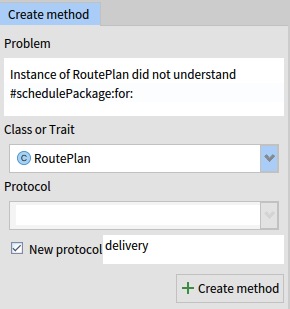}
\caption{The debugger proposes to create unknown methods.\label{debugger-create}}
\end{figure}

\paragraph{Building from the debugger.}
The process begins with writing a unit test before any implementation exists.  
For example, the developer specifies that a package located within a depot’s service area should appear in today’s delivery plan by calling something like \texttt{RoutePlan schedulePackage: aPackage for: aDate}.  
The developer executes the test that fails, as expected because the method \texttt{RoutePlan>>schedulePackage:for:} does not exist.
The debugger opens and prompts the developer to create that missing method in the \texttt{RoutePlan} class (Figure~\ref{debugger-create}).

\begin{figure}[H]
\includegraphics[width=0.8\linewidth]{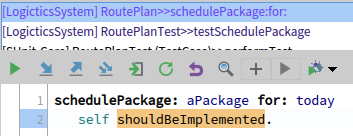}
\caption{The developer implements the missing method in the debugger and resumes the execution.\label{debugger-method-created}}
\end{figure}

The developer uses the debugger to automatically create the missing method.
The debugger creates this method and inserts a call to that method in the running execution, showing a new stack frame on that method (Figure~\ref{debugger-method-created}).
Instead of returning to a code editor, the developer implements the method \emph{inside the debugger itself}, 
 then resumes the paused execution and the test now passes---all within a single live debugging session.  

\paragraph{Generating code with live values.}
Let us imagine that the developer continued their development, and are now building new tests.
They wrote a method \texttt{RoutePlan>>defaultSchedulePlan} and they want to write first basic tests.
They would like to write simple tests such as \texttt{self assert: RoutePlan new defaultSchedulePlan equals: <something>} but they have yet to determine that \texttt{<something>}.
They have to either manually check in the source code or to execute an incomplete test with a breakpoint to observe the live value.
Instead, the Pharo debugger lets developers write test code such as \texttt{self try: RoutePlan new defaultSchedulePlan}.
Upon execution, that test opens a debugger that transform the \textit{try assertion} to a real assertion using run-time values, \textit{e.g.}, \texttt{self assert: RoutePlan new defaultSchedulePlan equals: 'success'}.

\begin{figure}[H]
\includegraphics[width=\columnwidth]{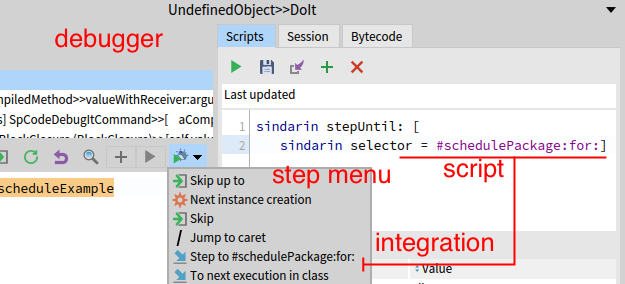}
\caption{Creation and integration of a new stepping tool.\label{custom-step}}
\end{figure}

\paragraph{Extending the debugger with domain-specific tools.} 
Now, the development of the logistic system has progressed and there are multiple scheduling strategies for delivery.
The developer wants to explore the execution of delivering strategies by using step-by-step execution in the \texttt{schedulePackage:for:} method.
However, since there are now multiple scheduling strategies, there are many methods with the same signature.
Therefore, it is hard to predict which implementation will be called, and it is impractical to set breakpoints in every one of them.
This scenario can be approached using \textit{moldable debugging} methods~\cite{Chis14b}, such as extending the debugger with domain-specific tools.
The Pharo debugger exposes the \textit{Sindarin} debugging API~\cite{Dupr19a} that developers can use to adapt the debugger to debug their problems.
For example, in Figure~\ref{custom-step} we see an open debugger with a focus on the scripting pane.
In there, we wrote a script that automatically steps the execution until a method with the \texttt{schedulePackage:for:} is executed.
The script can be executed directly or automatically integrated as a new stepping command in the general debugger toolbar.
In effect, we have created a custom stepping operator that advances execution until it reaches our scheduling method, then halts to return control to the developer.

\section{Scenario: Extending the IDE in action}
In the previous scenario, the developer implemented delivery scheduling logic.
As the system grows, understanding geographic data becomes essential: depots have service areas, packages are associated with locations, and routing decisions depend on boundaries and regions.
Working only with raw objects or XML structures quickly becomes cumbersome.
In Pharo, however, developers are not limited to textual or tabular representations of data—they can extend the environment itself to better match their domain.

In Pharo, the go-to tool for browsing any object is the inspector.
By default, an inspector will show the contents of the object's instance variables, but this behavior can be easily customized to show more useful information in different ways.
Objects can define their own custom views (tables, graphs, UI previews, domain-specific widgets), and these tools run in the same object space as the application.
This makes analysis and understanding a live, interactive activity instead of a static or external phase.

\begin{figure}
\includegraphics[width=0.5\linewidth]{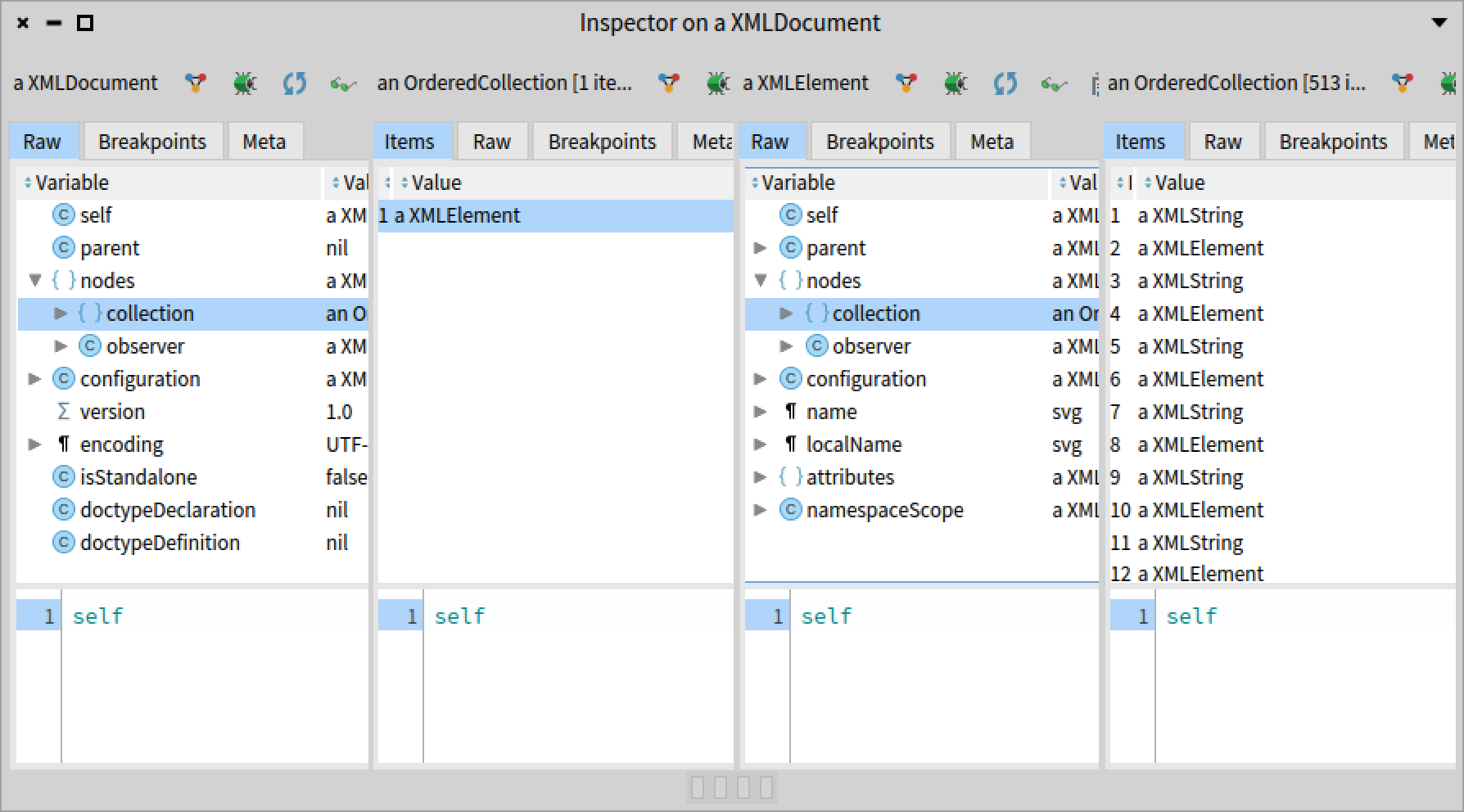}
\caption{Navigating the tree structure of an SVG world map file.\label{nav}}
\end{figure}

To illustrate this, imagine that the developer loads an SVG file representing a world map in order to define delivery regions.
After parsing the SVG with an \texttt{XMLDOMParser}, they may want to explore the internal structure of the document to locate countries or shapes.
This can be achieved by creating the document and opening it directly in an inspector:

\begin{lstlisting}
(XMLDOMParser parse: 'world.svg' asFileReference readStream contents) document inspect 
\end{lstlisting}

Developers can then freely navigate the tree structure, as shown in Figure~\ref{nav}.

After some simple processing, the developer produces a list of domain objects, each representing a country with its associated SVG path.
However, the default visualization is not very informative, since most objects present themselves only as instances of their class (e.g., \texttt{an EarthMapCountry}).
To improve readability, the first step is to redefine \texttt{printOn:} so that each object shows its country name throughout the IDE (e.g., \texttt{an EarthMapCountry (France)}).

\begin{figure}
\includegraphics[width=0.5\linewidth]{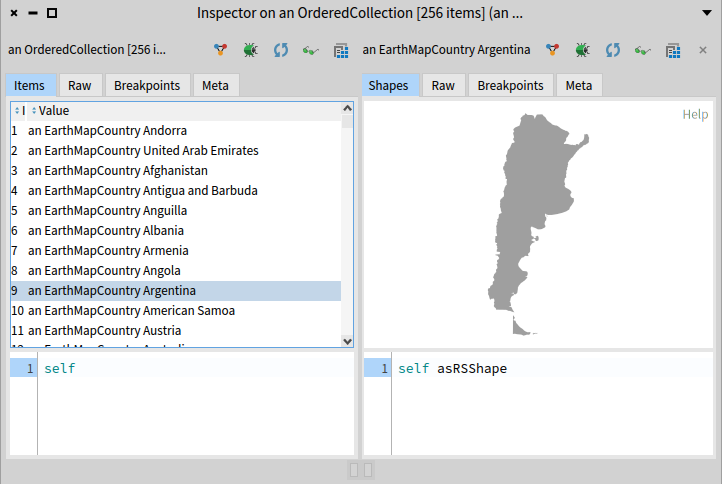}
\caption{Inspecting a list of country objects and seeing their visual aspects.\label{exte}}
\end{figure}

While this improves identification, a more expressive option is to extend the inspector itself to present a domain-specific view.
Listing~\ref{lst:insp} shows how to add a custom inspector tab that renders the actual shape of the country.
With this extension, developers can visually inspect each country directly inside the environment, as shown in Figure~\ref{exte}.

\begin{lstlisting}[label={lst:insp}]
EarthMapCountry >> inspectorShape: aBuilder
    <inspectorPresentationOrder: 0 title: 'Shape'>
    
    | canvas |
    canvas := RSCanvas new.
    canvas add: self asRSShape.
    canvas @ RSCanvasController.
    ^ SpRoassalInspectorPresenter new
        canvas: canvas;
        yourself
\end{lstlisting}

This shows how developers can not only inspect data, but adapt and extend the IDE to the domain at hand.
In this case, a geographic visualization becomes a first-class development tool, supporting understanding and decision-making in the live system.

\section{Scenario: Evolution}
Refactoring in Pharo is not limited to renaming: it  supports transformations and refactorings
\cite{Saren24a}. 
Because the environment is live and reflective, developers can deprecate a method, automatically track all usages, and apply context-sensitive rewrite rules to transform or migrate call sites incrementally.
This enables safe evolution over time, much harder to achieve in traditional file-based IDEs.

In our scenario, the delivery feature has been in use for a while.
The developer realizes that the method \sf{schedulePackage:for:} is poorly named.
Before changing it, they use Pharo’s visualization tools (e.g., senders/implementors browsers or Inspector-based views) to understand how widely the method is used and how invasive the update might be.
Since the system records every compilation action, the developer can roll back to previous microcommits and verify the effect of the change on the examples and tests.
Finally, rather than blindly renaming all occurrences of a method, the developer can mark the method as deprecated with automated rewriting behavior.
When the code is executed, the environment then automatically rewrites each executed call site \cite{Duca22a}.

\section{Discussion}

Although Pharo’s live object model enables these scenarios, many of the underlying ideas could inspire new directions in mainstream IDEs.
These environments are still organized around discrete phases (edit, build, run, debug) but the distance between them and live environments could be bridged.
Modern IDEs do offer mechanisms that could evolve toward similar practices: hot-reload in the JVM, inline test execution in VS Code, and stateful notebooks in JetBrains Fleet all point to continuity between coding and execution.

Xtreme TDD and Debugger-driven development, for instance, could become more integral if debuggers supported in-context code creation rather than post-hoc inspection.
Instead of opening a separate editor, a user could define or refine a function directly in a paused frame and resume execution seamlessly.
Similarly, domain-specific inspectors could be realized through existing plugin architectures.
JetBrains already provides structured views (e.g., database table viewer or JSON tree explorer), but these are tied to file formats rather than live domain objects.
Opening the plugin API to let developers craft custom, object-aware visualizations could enrich environments with a moldable inspection model.

Even refactoring and evolution mechanisms can benefit from a live perspective \cite{Teso20a}.
Tools like IntelliJ’s structural search and replace or VS Code’s semantic refactoring already reason about syntax trees; extending them to operate on live execution contexts, with awareness of dynamic call sites, would blur the boundary between static code and running system.
These examples suggest that live-environment ideas do not depend on reflection or image-based systems—they can emerge gradually within file-based ecosystems.

Adopting such features would not merely enrich developer experience but could influence how development workflows are structured.
Live, object-centric feedback loops foster experimentation, faster validation, and a more conversational relationship with the program under construction.
Using Pharo as inspiration thus highlights an opportunity: to rethink how mainstream IDEs might close the gap between editing and running, turning development from a staged pipeline into an ongoing dialogue with the system itself.

Scalability deserves special consideration, as liveness introduces challenges like global state management and team coordination.
While there are large Smalltalk systems in the industry, this is an opportunity for discussion on how IDE design choices support system scale.

\section{Conclusions}
Traditional IDEs have evolved greatly since their origins, and helped shape and speed up the development process.
Despite this progress, the underlying metaphor has not changed: development is still framed as a sequence of file manipulations rather than a dialogue with a live system.
In this paper, we presented specific features of Pharo, a live coding environment, not just for advertising it, but inviting reflection on how an IDE with a different and in some ways unique history could help rethink the usual toolset and even workflow.
Simply stepping outside the inherited model of file-based editing could make room for features that make development more fluid, exploratory, and responsive to the system as it runs.

\begin{acks}
We thank the Pharo consortium and the community for all their contributions to Pharo. Authors acknowledge grant PICT-2019-02485 from Agencia I+D+i.
\end{acks}

\bibliographystyle{ACM-Reference-Format}
\bibliography{others,rmod}

\end{document}